\documentclass[12pt,preprint]{aastex}

\begin{document}

\title {THE SURVIVAL  OF  WATER 	WITHIN EXTRASOLAR MINOR PLANETS}

\author{M. Jura\altaffilmark{a}, and S. Xu\altaffilmark{a,b}}

\altaffiltext{a}{Department of Physics and Astronomy, University of California, Los Angeles CA 90095-1562; jura@astro.ucla.edu; xsynju@gmail.com}
\altaffiltext{b}{English translation of \includegraphics[width=1.2cm]{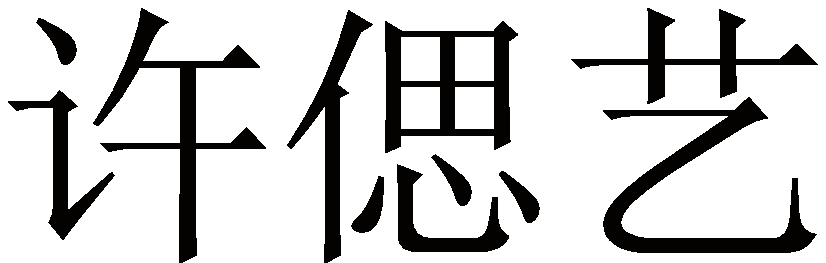}}

\begin{abstract}
We compute that extrasolar minor planets can retain much of their internal H$_{2}$O during their host star's red giant evolution.  The eventual accretion of a water-rich body or bodies onto a helium white dwarf might supply  an observable amount of
 atmospheric hydrogen, as seems likely for GD 362.  More generally, if hydrogen pollution in helium white dwarfs typically results from accretion of large parent bodies rather than interstellar gas as previously supposed, then H$_{2}$O  probably constitutes  at least 10\%  of the aggregate mass of  extrasolar minor planets.  One observational test of this possibility is to examine the atmospheres of externally-polluted white dwarfs  for  oxygen in excess of that likely contributed by oxides such as SiO$_{2}$.  The relatively high oxygen abundance previously reported in
 GD 378 plausibly but not uniquely can be explained by accretion of an H$_{2}$O-rich parent body or bodies.   Future ultraviolet observations of white dwarf pollutions can serve to investigate the hypothesis  that environments with liquid water  that are suitable habitats for extremophiles are  widespread in the Milky Way.
\end{abstract}
\keywords{planetary systems -- stars, white dwarf}

\section{INTRODUCTION}

Material accreted onto white dwarf atmospheres from orbiting disks created from tidally-disrupted parent bodies can be used as a powerful tool to measure  the bulk composition of extrasolar asteroids or minor planets  (Jura 2008)\footnote{The distinction, if any, between an asteroid and a minor planet is vexing (Hughes \& Marsden 2007).  Our  informal sense is
that an asteroid may have essentially any size while a minor planet is a ``larger" object.  There may be a physical difference; for example, the larger
objects may experience differentiation.  Here, we make the arbitrary choice that objects smaller than 100 km in radius are denoted as asteroids while objects larger than this
radius can be described either as asteroids or minor planets.  We accept  this uncomfortable ambiguity for the classification of the larger objects because of historical usage.  For example, Ceres, with a radius near 500 km,
is usually described as an  asteroid.}.  In this paper we focus on using white dwarfs as tools for studying H$_{2}$O within extrasolar planetary systems.

Water may be  a major constituent of extrasolar asteroids and planets.  For example, in the Solar System, Ceres, the most massive asteroid, has a mean density of 2.1 g cm$^{-3}$ suggesting ${\sim}$25\% of its mass is water (McCord \& Sotin 2005, Thomas et al. 2005).  With a mass of 9.4 ${\times}$ 10$^{23}$ g (Michalak 2000), Ceres by itself possesses ${\sim}$25\% of the mass of the entire asteroid belt which totals to  ${\sim}$3.6 ${\times}$ 10$^{24}$ g (Krasinksy et al. 2002).   Therefore, at least 6\% and quite possibly more (Jewitt et al. 2007) of the matter in  the Solar System's asteroid belt is water.  Beyond  the asteroid belt, there are numerous ice-rich objects  such as  Callisto (Canup \& Ward 2002), the outermost Galilean satellite of Jupiter.    If extrasolar planetesimals form beyond a ``snow line" (Sasselov \& Lecar 2000), they also may possess  an appreciable fraction of water. Such water within extrasolar asteroids may be important in the history and evolution of a system's analogs to terrestrial planets.  For example, it is likely that asteroids delivered the bulk of the Earth's water during the evolution of the early Solar System (Morbidelli et al. 2000), and, by analogy, a similar process could occur elsewhere leading
to extrasolar oceans.  Environments with  liquid water  may support extremophile life  (Rothschild \& Mancinelli 2001).

A star's luminosity during its red giant evolution is sufficiently high that icy comets within 40 AU of the host star are sufficiently heated that they are fully sublimated and destroyed (Jura 2004).  Similarly, we expect that surface ice on an  extrasolar asteroid would be lost.  However, as computed below in ${\S2}$, internal water may never be vaporized and therefore
survive within a sufficiently large asteroid, even during the host star's high luminosity asymptotic giant branch (AGB) evolution.  If this minor planet's orbit is sufficiently perturbed that it is tidally-disrupted and subsequently accreted onto the star during its white dwarf phase (Debes \& Sigurdsson 2002, Jura 2003), the resulting  pollution of hydrogen and oxygen may be detectable.

Approximately 20\% of white dwarfs in the temperature range between 8000 K and 20,000 K, a class of stars with cooling ages greater than 5 ${\times}$ 10$^{7}$ yr,  have atmospheres where helium is the dominant element (Tremblay \& Bergeron 2008).  While elements heavier than helium diffuse below the outer convective zone  in less than 10$^{6}$ yr (Koester 2009), hydrogen is lighter
than helium and remains in the star's outer mixing zone during the entire white dwarf phase of evolution. Two recent surveys have found that the mass of hydrogen in the atmospheres of helium white dwarfs increases with the cooling age of the star (Dufour et al. 2007, Voss et al. 2007),  suggesting ongoing hydrogen accretion in these stars.

The interstellar medium has usually been assumed to be the source of the hydrogen accreted onto helium white dwarfs (see, for example, MacDonald \& Vennes 1991).  However, the importance of this process is  uncertain.  Koester (1976) argued that  Bondi-Hoyle theory vastly overestimates the  accretion rate since the mean free path
between interstellar atoms is greater than the typical accretion radius and the fluid approximation fails.  In contrast, Alcock \& Illarionov (1980)
argued that for an ionized plasma, the fluid rate is appropriate, although this is uncertain, especially if the white dwarf also is
magnetized (Wesemael \& Truran 1982).  Interstellar accretion  may also be suppressed if a white dwarf has a weak wind (MacDonald 1992).  Therefore, alternative models for the accretion of hydrogen should be considered.

Jura et al. (2009) proposed that the unusually large amount of 7 ${\times}$ 10$^{24}$ g of hydrogen in GD 362, a helium white dwarf can be understood
if we are witnessing the aftereffects of the destruction of a  parent body with internal ice and a mass somewhere between Callisto's and  Mars's.   Here, we explore whether other white dwarfs  may have acquired their much smaller masses  of atmospheric hydrogen from minor planets with internal water.

In {\S2}, we present our model for when internal water is retained by a minor planet during its host  star's red giant evolution.   In {\S3} we present observational tests of the models.  We consider both the specific examples of GD 362 and GD 378, and a more general assessment of the model that extrasolar minor planets
retain water that ultimately is detectable by hydrogen and/or oxygen pollution of white dwarfs.   In {\S4} we put our results into a broader context and  summarize our conclusions.

\section{SURVIVAL OF A MINOR PLANET'S INTERNAL WATER DURING ITS HOST STAR'S RED GIANT EVOLUTION}
The survival of rocky asteroids during a star's AGB evolution has been discussed by Jura (2008); the most important
pathway for their destruction appears to be thermal sublimation.   In these calculations, asteroids composed of olivine with radii greater than 50 km which initially orbit at  least 2.5  AU from  their host star are likely to survive the  AGB evolution.  In this paper, we focus on asteroids with initial orbits of 5 AU or larger and which therefore are cool enough to have
large amounts of internal water.  We consider rocky objects without atmospheres so that ultraviolet heating of the exosphere described by Villaver  \& Livio (2007) that leads to the erosion of gas giants is not important in the models computed here.

During a star's high luminosity phase, an asteroid's surface temperature can exceed 400 K and surface water would sublimate
and presumably be lost.  However, the interior of an asteroid does not attain the surface temperature because the time required for the inward conduction of
heat can be longer than the duration of the star's AGB phase.
Here, we describe a simple model to describe the internal temperature as a function of time and radius and therefore compute the times and locations
when and where water becomes gaseous.  We further assume that if gaseous, the water is quickly vented and lost from
the minor planet.

Following the calculations by Ghosh \& McSween (1998) for Vesta, we assume that the temperature, $T(R,t)$, within a spherically symmetric extrasolar asteroid is governed by thermal conduction:
\begin{equation}
\frac{{\partial}T}{{\partial}t}\;=\;\frac{1}{R^{2}}\frac{{\partial}}{{\partial}R}\left(R^{2}\,{\kappa}\,\frac{{\partial}T}{{\partial}R}\right)
\end{equation}
where ${\kappa}$ is the thermal diffusivity. Following Turcotte \& Schubert (2002), we take ${\kappa}$ = 10$^{-2}$ cm$^{2}$ s$^{-1}$ as representative of terrestrial rocks and presumably also of extraterrestrial asteroids.  The thermal diffusivity of ice is less than this value, and
therefore our calculations may overestimate the rate at which heat is conducted into the interior of a minor planet and thus overestimate
the amount of water that is lost.
As with the previous models for Vesta (Ghosh \& McSween 1998), we assume that radioactive heating in the interior of the asteroid is negligible.

The solution to Equation (1) depends upon the boundary condition.
 If the albedo is negligible, the  temperature at the outer radius of the minor planet, $R_{0}$, is:
\begin{equation}
T(R_{0},t)\;=\;\left(\frac{L_{*}(t)}{16\,{\pi}\,{\sigma}_{SB}\,D^{2}(t)}\right)^{1/4}
\end{equation}
where $L_{*}(t)$ is the luminosity of the star as a red giant and $D(t)$ is the distance of the asteroid from its host star.
At  the onset of the star's red giant evolution,  because  radioactive heating is likely to be unimportant, the minor planet's internal temperature is taken as uniform and
equal to the surface temperature determined from Equation (2).
We assume that $R_{0}$ is constant during the star's AGB evolution because the shrinkage by sublimation of a surface assumed to be composed of olivine is less than
10 km for the conditions of interest (Jura 2008).

We use the solar models of Girardi et al. (2000) and Pols et al. (1998) for $L_{*}(t)$ for stars with main-sequence masses of 3 M$_{\odot}$ and 1  M$_{\odot}$ and $Z$ = 0.019, respectively.  We choose these two masses because they span the range of the main-sequence progenitors of most
white dwarfs.
The results for $L_{*}(t)$ are displayed in Figures 1 and 2.

 Following Schroder \& Cuntz (2005, 2007), we assume that the host
star loses mass with the rate, $dM_{*}/dt$ governed by the expression:
\begin{equation}
dM_{*}/dt\;=\;{\eta}\,\left(\frac{L_{*}(t)}{L_{\odot}}\right)\left(\frac{R}{R_{\odot}}\right)\left(\frac{M_{\odot}}{M_{*}}\right)\left(\frac{T_{*}}{4000}\right)^{3.5}\,\left(1\,+\,\frac{g_{\odot}}{4300\,g_{*}}\right)
\end{equation}
where $g_{*}$ is the gravity of the star and ${\eta}$ = 8.0 ${\times}$ 10$^{-14}$ M$_{\odot}$ yr$^{-1}$.  We truncate the AGB evolution
when the star's mass shrinks to 0.55 M$_{\odot}$ and 0.68 M$_{\odot}$ for stars with main-sequence masses of 1.0 M$_{\odot}$ and 3.0 M$_{\odot}$, respectively (Weidemann 2000)\footnote{The mass loss from the star is not treated self-consistently in the sense that Equation (3) is not used in the stellar evolutionary calculations but only for
the evolution of the minor planet's  orbit  described below.  Pols et al. (1998) note that the star's evolution on the AGB is not especially sensitive to the mass loss rate; Iben \& Renzini (1983) state that an AGB star's luminosity is mainly determined by its core mass.}

The orbital distance  of the minor planet from the host star either can decrease because of wind drag or increase because the star loses mass
and the asteroid becomes less tightly gravitationally bound.
 Following Jura (2008), an asteroid with a radius greater than 3 km at an orbital separation greater than 3 AU encounters less than its own
 mass in the wind and drag does not completely dominate the asteroid's orbital evolution.  Since we mostly consider minor planets with radii much larger than 100 km and initial orbital radii of at least 5 AU,  we neglect drag's effect on the object's orbit.  As a result, we assume that the minor planet orbits with  constant angular momentum.
Therefore, as the host star loses mass, then:
\begin{equation}
\frac{D(t)}{D(0)}\;=\;\frac{M_{*}(0)}{M_{*}(t)}
\end{equation}
We employ $D(t)$ from Equation (4) when we compute the object's surface temperature with Equation (2).

The minor planet's internal pressure, $p$, is derived with the assumptions that it  is in hydrostatic equilibrium with uniform density, ${\rho}_{0}$, and zero surface pressure.  Thus (Turcotte \& Schubert 2002):
\begin{equation}
p(R)\;=\;\frac{2{\pi}}{3}\,{\rho}_{0}^{2}\,G\,\left(R_{0}^{2}\,-\,R^{2}\right)
\end{equation}
Water is assumed gaseous if  the temperature exceeds the vaporization temperature for the local pressure $p(R)$.  Fitting   the phase diagram for water  of Keenan et al. (1969) to within 2\% for the pressures of interest and using cgs units,  the water  is vaporized  when:
\begin{equation}
T\;{\geq}\;78.45\,\ln\left(\frac{p\,+\,6554000}{60640}\right)
\end{equation}
 In all our models, we assume the minor planet has a density of 2.1 g cm$^{-3}$, similar to the value for Ceres. If the density is greater than this value, then,
 by Equation (5), the internal pressure would be larger and consequently, by Equation (6), the vaporization temperature also would be larger.  If so, less water would be lost then we compute below.

We first consider a model that could be applied to GD 362 which probably had a main-sequence progenitor near 3 M$_{\odot}$ (Kilic et al. 2008).  The temperature as a function of internal radius at different times is shown in Figure 3 for a minor planet of radius 100 km at an initial orbital separation of 5 AU.  We see that the temperature at the center of the asteroid is unchanged with time and equal to that given by the initial conditions.   Also, initially, the temperature is low enough that water can be retained throughout the asteroid. By the end of  the AGB phase, water in the outer 7 km is vaporized.  As described by Turcotte \& Schubert (2002) and as  can be derived from our Equation (1) by dimensional analysis,
in time ${\tau}$, the characteristic distance that a thermal pulse can propagate is $({\tau}\,{\kappa})^{1/2}$.  As shown in Figures 1 and 2,  the time interval  of the star's high luminosity AGB phase is  ${\sim}$ 6 ${\times}$ 10$^{6}$ yr, and therefore, for  our adopted value of the thermal diffusivity, the temperature would be  elevated over a depth of
${\sim}$ 14 km. This qualitative estimate of the depth of the heating zone is consistent with the more exact calculation.  A notable feature of the model shown in Figure 3 is that at the end of the AGB evolution, the maximum  temperature occurs below the minor planet's surface.  This temperature ``inversion" is a result of the minor planet's orbital separation increasing with time and therefore the surface becoming relatively cool. In this phase, the interior temperature lags behind the fall of the surface temperature.  In general, because the heating history of the asteroid is complicated, the internal temperature profiles can be somewhat complex.

 We show in Figure 4 the mass percentage of ice that is retained in asteroids of an initial radius between 10 km and 200 km and at initial orbital separations of 5 AU, 7.5 AU and 10 AU.  As expected,
the asteroids at greater distances from the star retain more water and larger asteroids retain a greater percentage of their initial water.  For all three cases, asteroids
with a radius of at least 100 km retain at least half of their internal water, and even much smaller asteroids can retain some ice.

We next consider a model for a star of 1 M$_{\odot}$ that could be applied to the future evolution of the Sun.  In Figure 5, we show the temperature profiles with time of a minor planet with a radius of 100 km at an orbital distance of 5 AU.    The minor planet begins its evolution at a much lower
temperature than for the 3 M$_{\odot}$ star, but the duration of the heating event is longer.  Until the last 7 ${\times}$ 10$^{8}$ yr, the internal temperature of
the minor planet hardly changes.  During the last phases of the evolution, there is a modest rise in temperature even in the interior of the minor planet.  However, the
 large temperature increase required to vaporize the water only occurs in the  outer ${\sim}$10 km.  Finally, in Figure 6, we display the results for
 the percentage of water that is retained for asteroids of radii ranging from 10 km to 200 km and initial orbital separations of 5 AU, 7.5 AU and 10 AU.  There is more
 effective
 retention than for
 the star with a main sequence mass of 3 M$_{\odot}$.  For example,  we see that in all three cases, at least 70\% of the internal water is retained for asteroids of radius of 100 km.

 We conclude that minor planets  can maintain much of their internal water in
 a wide variety of circumstances.  Therefore,  minor planets ultimately might deliver appreciable amounts of hydrogen and oxygen to white dwarfs.

\section{OBSERVATIONAL CONSEQUENCES}
 We now
consider observational tests of the model that extrasolar asteroids possess internal water.
\subsection{Hydrogen}

Jura et al. (2009) proposed that the hydrogen in the outer mixing zone of GD 362, a helium white dwarf,
may have resulted from the accretion of a a parent body at least as massive as Callisto which has a radius of 2400 km\footnote{There is a typographical error in the Abstract of Jura et al. (2009) where the approximate mass of hydrogen in GD 362 is given as 0.01 M$_{\oplus}$ instead of the correct value of 0.001 M$_{\oplus}$.}.  Since much water can survive in minor planets with radii of 100 km, our calculations
indicate that almost all of the internal water in the parent body or bodies responsible for the accretion onto GD 362 could have survived the AGB phase of the star's pre-white-dwarf evolution.

Beyond the special  case of GD 362 with its very large amount of atmospheric hydrogen, we now consider whether it is  possible that the more ordinary amounts of hydrogen in helium white dwarfs are derived from minor planets.
We show in Figure 7 a plot of the mass of hydrogen in the outer mixing zone of helium white dwarfs in the cooling age range between 0.1 and 1.0 Gyr from Voss et al. (2007) and Dufour et al. (2007).  The cooling ages are estimated by Voss et al. (2007); for the other stars we estimate cooling agres from the star's effective temperature as in Farihi et al. (2009).  We do not consider stars with  cooling ages less than 0.1 Gyr because it is possible
that a stellar wind can suppress accretion (MacDonald 1992).  We do not consider white dwarfs with cooling ages greater than 1.0 Gyr because at some time beyond this age,
the mixing layers of stars that have hydrogen envelopes can become sufficiently massive that interior helium is dredged-up to the surface and
these stars can appear as helium objects (Tremblay \& Bergeron 2007).  In these cases,  the atmospheric hydrogen is not the result of accretion.   In Figure 7, we also show
the expected mass of atmospheric hydrogen if the star accretes with a rate of 6 ${\times}$ 10$^{5}$ g s$^{-1}$, the average from the Voss et al. (2007) survey if we assume that
the stars with upper limits to their atmospheric hydrogen are not accreting material.

We see in Figure 7 that the average of the  mass accretion
rate passes through the values reported by Voss et al. (2007) but below the values reported by Dufour et al. (2007).  The stars identified by Dufour et al. (2007) were taken from the Sloan Digital Sky Survey; many helium white dwarfs were observed and only the ones with the strongest hydrogen lines were recognized.  The survey
by Voss et al. (2007) was more sensitive at detecting hydrogen in individual stars since it used a larger telescope and  higher spectral resolution.      It is therefore not too surprising that the line denoting the mean accretion rate falls below the points from the sample of Dufour et al. (2007).

To assess further the hypothesis that the hydrogen accretion into helium white dwarfs is derived from minor planets,  we now compare the mean hydrogen accretion rate with the mean accretion rate of heavy atoms.  Because a comprehensive treatment of the accretion onto the entire population of white dwarfs is not available, we must
extrapolate from the current imperfect studies.

 Koester \& Wilken (2006) report accretion rates for 38 polluted hydrogen white dwarfs, 8 with an infrared excess, 24 observed with the {\it Spitzer Space Telescope} and found not to have an infrared excess and 6 which have not been observed with the {\it Spitzer Space Telescope} and it is not known if they have an excess.  The stars with an infrared excess almost certainly are accreting from a disk created by a tidally-disrupted asteroid or minor planet (see von Hippel et al. 2007, Kilic \& Redfield 2007, Jura 2008). For the stars without an excess, it
is not known whether their accretion is from asteroids or from the interstellar medium.
 For the 8 stars with an infrared excess (Farihi et al. 2009, 2010a) and with the assumption that
the heavy element accretion rate is 0.01 of the accretion rate given by Koester \& Wilken (2006) which presumed a large amount of hydrogen, the mean value of the heavy atom accretion rate  is 9.5 ${\times}$ 10$^{8}$ g s$^{-1}$.   The average accretion rate for the stars without an infrared excess is about 1/3 of this value.  Because in this sample there are three times as many stars that accrete and do not have an infrared excess, it appears that in aggregate, white dwarfs accrete approximately as much matter when they
do not have an infrared excess as when they do.
Farihi et al. (2009) found  that between 1\% and 3\% of white dwarfs with cooling ages less than 0.5 Gyr have an infrared
excess.  Therefore, to estimate the lower bound of the mass accreted by the entire sample,   we assume that at any given moment,   only 1\% of white dwarfs have infrared disks and these are the only systems where there is asteroidal accretion.  In this
case, the average heavy atom accretion rate from asteroids onto all white dwarfs is 9.5 ${\times}$ 10$^{6}$ g s$^{-1}$. To estimate the upper bound to the mass accretion rate, we assume that  3\% of all white dwarfs
have infrared dust disks and that as much mass is delivered by asteroids when the star does not have an infrared excess as in the epochs when the star
does have an excess.  With these assumptions, the mean heavy atom accretion rate is 6 ${\times}$ 10$^{7}$ g s$^{-1}$.

To estimate the minimum fraction of ice in
the parent bodies, we consider the maximum average heavy element accretion rate.  Since hydrogen provides 0.11 of the mass of H$_{2}$O, then at least ${\sim}$ 10\% of the mass
in the minor planets is water if these objects deliver the bulk of the hydrogen in helium white dwarfs.  Also, if this scenario accounts for much of the  hydrogen in helium
white dwarfs and  since at least 55\% of helium white dwarfs possess atmospheric hydrogen (Voss et al. 2007), then over half of their progenitor main-sequence stars must have had planetary systems.

\subsection{Oxygen}
 In  bulk Earth and chondrites, the dominant elements are oxygen, silicon,
magnesium and iron (Allegre et al. 1995, Wasson \& Kallemeyn 1988), and to-date, a similar pattern appears to hold in extrasolar minor planets (Zuckerman et al. 2007, Klein et
al. 2010).  If a white dwarf has accreted a water-rich minor planet, it may display an ``excess" amount of  oxygen over what would be contributed by
rocky minerals.

 Ultraviolet spectra from the FUSE satellite were used to identify oxygen in GD 61 and GD 378 (Desharnais et al. 2008), while optical spectra obtained with the HIRES echelle spectrograph at Keck Observatory have been used to identify oxygen in GD 40 (Klein et al. 2010)\footnote{Provencal et al. (2005) have
discovered oxygen and carbon emission is two white dwarfs with $T_{eff}$ near 12000 K which may be the result of chromospheric
activity.  They report only an upper limit to the amount of photospheric oxygen.  In these unusually massive white dwarfs, the atmospheric oxygen likely was dredged-up from the interior in
these helium stars and not accreted from an external source.}.  In the case of GD 40, there is enough  magnesium, silicon, calcium  and iron,  that all the oxygen in the parent body  could have been bound into the oxides MgO, SiO$_{2}$,
CaO, FeO and Fe$_{2}$O$_{3}$
(Klein et al. 2010).  Therefore, there is no evidence for water in the parent body accreted onto GD 40.    As discussed in detail below, GD 61 and GD 378 nominally  possess
``excess" oxygen, although we argue that only in GD 378's case does the evidence support accretion of a water-rich parent body or bodies.

We show in Table 1 the masses of pollutants in the outer mixing layers of both GD 61 and GD 378 by using both the relative  abundances derived by Desharnais et al. (2007) and  calculations by Koester (2009, private communication\footnote{These results are extensions  of the calculations by Koester (2009) with the specific  temperatures, gravities and hydrogen to helium ratios determined by Desharnais et al. (2008) for these two stars.}) for the total masses of the mixing layers for these two stars.   Since
oxygen is  the most abundant element heavier than helium, these two stars  are candidates for having accreted from
ice-rich parent bodies.

GD 378 does not have an infrared excess (Mullally et al. 2007) while GD 61 has not been observed with the {\it Spitzer Space Telescope}, and it is not currently known whether it has an infrared excess.  It is therefore possible that both stars have experienced accretion from the interstellar medium rather than from tidally-disrupted asteroids; we now evaluate this hypothesis.
In GD 61, GD 378, and the Sun, $m$(C)/$m$(O), the relative abundance by mass, is $<$ 8 ${\times}$ 10$^{-4}$, 0.092, and 0.38, respectively\footnote{Here we follow geophysical rather than astrophysical convention and report abundances by mass rather than by number.}  (Desharnais et al. 2007, Lodders 2003).  Therefore,  carbon is underabundant relative to the Sun in these two white dwarfs, and it is unlikely that they are polluted by accretion of interstellar matter\footnote{As listed in Table 1, the carbon settling time is longer than the oxygen settling time so that $m$(C)/$m$(O)  in the parent body must be less than or equal to the ratio in the
star's photosphere.  Therefore, the true carbon deficiency may be more marked than these values.  On the other hand, the carbon ``deficiency" in GD 378 is ``only" about a factor of 4. Since Desharnais et al. (2007) report errors in the carbon and oxygen abundance of factors of 1.25 and 2, respectively, their data do not strongly rule out
interstellar accretion for GD 378.  GD 61 almost certainly has experienced asteroidal accretion.}.   Also, in GD 61, GD 378,  and the Sun $m$(H)/$m$(O) equals 4.2, 3.5, and  130, respectively (Desharnais et al. 2007, Lodders 2003).  Therefore, even though hydrogen does not settle below the mixing layer, it is underabundant relative to oxygen when compared with the
likely interstellar value by more than a factor of 10, an additional argument against interstellar accretion.  However, this argument is not conclusive since interstellar accretion of hydrogen relative to  heavy elements may be very inefficient (Dufour et al. 2007, Voss et al. 2007).  We conclude that the cases are moderately strong and very strong,
respectively, that GD 378 and GD 61 have experienced asteroidal accretion.

From Table 1, the total mass of detected heavy elements in the mixing layers for GD 61 and GD 378  is 1.7 ${\times}$ 10$^{21}$ and
3.8 ${\times}$ 10$^{21}$ g, respectively.  Assuming that the accreted matter had a mean density of 3 g cm$^{-3}$, then the minimum radii of
the parent bodies for GD 61 and GD 378 were 51 and 67 km, respectively if polluted by a single object\footnote{These radii are conservatively low, and would be  larger if we assumed a density of 2.1 g cm$^{-3}$ as in our calculations in ${\S2}$}.  According to Figures 4 and 6,   parent bodies of this size  can retain appreciable initial  internal water.
Even if they have been recently impacted by multiple smaller bodies with radii near, say, 30 km, they may have accreted some water.

We now estimate the maximum fraction of
oxygen that could have been contained within rocky minerals in the parent bodies that accreted onto GD 61 and GD 378.   Desharnais et al. (2007) report oxygen, iron and
silicon abundances but not magnesium, an important oxide-bearing element in the Earth and chondrites.  With the assumption that silicon and iron were largely in the form SiO$_{2}$ and FeO in the parent body\footnote{In the Earth's crust, about 2/3 of the iron is contained within FeO and 1/3 within Fe$_{2}$O$_{3}$ (Ronov \& Yaroshevsky 1969). Therefore, the correction for iron in Fe$_{2}$O$_{3}$ rather than FeO is likely to be small, and we ignore this possibility.}, then from the results given
in Table 1, we  expect  there would be 3.5 ${\times}$ 10$^{20}$ g of oxygen in GD 61's mixing layer.   However, the observed value is
13 ${\times}$ 10$^{20}$ g, and therefore, there appears to
be some ``extra" oxygen.  Unless  magnesium or some other element is extremely overabundant compared to rocky material in the Solar System, the simplest interpretation of the evidence raises the possibility that a substantial amount of water may have been present in the parent body.  Applying the same assumptions to
the mixing layer for GD 378, we would expect 4.5 ${\times}$ 10$^{20}$ g of oxygen; the measured value is 26 ${\times}$ 10$^{20}$ g.  Again,
the evidence allows for  water in the parent body.

The simple approach for estimating relative abundances in the parent body adopted in the previous paragraph may not be realistic.
Since oxygen lingers longer than silicon and iron in the outer mixing layer, the current abundances in the mixing layer may not equal
the true abundances in the parent body.
Following the scenarios of
Koester (2009) and Jura et al. (2009) for the time evolution of helium white dwarf pollution, we assume a disk is very quickly formed from a
tidally-disrupted minor planet.  Material from this disk then accretes onto the star with an exponentially decaying rate with a characteristic time, $t_{disk}$.  As in the notation of Jura et al. (2009), we define a time parameter, ${\tau}(Z)$, for each element $Z$, such that
\begin{equation}
{\tau}(Z)\;=\;\frac{t_{disk}\,t_{set}(Z)}{t_{disk}\,-\,t_{set}(Z)}
\end{equation}
where  $t_{set}(Z)$ is the settling time in the mixing zone of element $Z$.  If  $t$ = 0 defines the  onset  of an accretion event when the disk is quickly formed and $M_{mix}(Z)$ is the currently measure mass of element $Z$ in the mixing layer, then the mass
of  element $Z$ in the parent body, $M_{par}(Z)$ is:
\begin{equation}
M_{par}(Z)\;=\;\frac{M_{mix}(Z)\;t_{disk}\;e^{t/t_{disk}}}{{\tau}(Z)\,(1\,-\,e^{-t/{\tau}(Z)})}
\end{equation}

 Since GD 378 does not display an infrared excess, this star
may be in a phase where the disk has largely dissipated ($t$ $>>$ $t_{disk}$) yet there is residual material in the mixing layer.  For the
case where $t$ $>>$ $t_{set}$ $>>$ $t_{disk}$,  Equation (8) becomes:
\begin{equation}
M_{par}(Z)\;{\approx}\;M_{mix}(Z)\,e^{t/t_{set}(Z)}
\end{equation}
Since iron and silicon settle more rapidly than oxygen, then perhaps the oxygen ``excess" is only a consequence of observing the system
in a late decay phase.   However, in this case,
the parent body may have been deficient in silicon compared to iron.  From Table 1, we see that currently in GD 378's atmosphere, the silicon to iron mass ratio is 0.62, essentially equal to  the bulk Earth value of 0.61 (Allegre et al. 1995).  If the system is a phase where we are witnessing  lingering aftereffects of  disk accretion, then the silicon to iron mass ratio in the parent
body would be less than the value in  bulk Earth since iron settles more rapidly than silicon.
For example, if we apply Equation (8) for the case that $t$ = 3 ${\times}$ 10$^{5}$ yr, then we find that the parent body masses
of oxygen, silicon and iron are 7.1 ${\times}$ 10$^{21}$, 1.2 ${\times}$ 10$^{21}$ and 6.9 ${\times}$ 10$^{21}$ g, respectively.  If so,
the inferred mass ratio of iron to oxygen is 1.0, comparable to the  bulk Earth value of 0.87 (Allegre et al. 1995).  A weakness of this particular model  is that   the silicon to oxygen
mass ratio of only 0.17 is  smaller than the value in the bulk Earth of 0.53.  We do not know enough about the composition of
extrasolar minor planets to rule out this possibility.  However,  because it requires an unfamiliarly low silicon abundance, this model seems slightly disfavored.

Another possibility is that GD 378 has recently experienced at least two impacts.  The first may have occurred so far in the past that only its oxygen lingers.  This would
be another way in which the apparent ``excess" oxygen abundance does not require a water-rich parent body.

 A further  hint to the composition of the parent body accreted onto GD 378
is that the star's photosphere  contains appreciable amounts of sulfur and, compared to other  polluted white dwarfs, a relatively large abundance of
carbon (Desharnais et al. 2007, Jura 2006).  Both carbon and sulfur are effectively volatile (Lodders 2003), and their relatively high abundance is consistent with the scenario that the parent body contained a substantial amount of  water.
We conclude that the evidence plausibly supports, but does not require, a scenario
where a water-rich parent body has been accreted by GD 378.

 In the case of GD 61, if we assume Equation (9) with $t$ = 1.5 ${\times}$ 10$^{5}$ yr  and the settling times in Table 1,
the mass ratios in the parent body of silicon and iron to oxygen are 0.45 and 0.80, respectively.  These ratios are close
to the bulk Earth values of 0.53 and 0.87, respectively (Allegre et al. 1995).  In this case, there is no clear ``excess" oxygen and no reason to argue  that water was present in
the asteroid accreted onto GD 61.

\section{CONTEXT AND CONCLUSIONS}

There is a huge range in possible water fractions within extrasolar asteroids and minor planets.  As much as half the
mass could be water (Leger et al. 2004), or, alternatively, there may be essentially none.  Currently, evidence is sparse.    Reach et al. (2009)  suggested that the infrared spectrum of
the white dwarf G29-38 can be best fit if there is water in the circumstellar disk.  However, this argument is model-dependent and not
certain.  Jura et al. (2009) have proposed that GD 362 has accreted from a water-rich object perhaps as massive as Mars.  Here, we argue there is plausible but not compelling evidence that there were significant amounts of water in the parent body or bodies that accreted onto GD 378.  Beyond these two specific polluted white dwarfs, the hydrogen abundances in helium white dwarfs might be explained with the unproven hypothesis that  at least  ${\sim}$10\% of the aggregate mass of extrasolar minor planets is water.

In ${\S3.2}$, we considered helium white dwarfs where photospheric oxygen has been detected.  One ambiguity in interpreting the
data is that the evolutionary phase of the accretion event is unknown, and the photospheric abundances may not represent the
true abundances in the parent body.  However, in hydrogen white dwarfs with $T$ $>$ 12000 K, the settling time of heavy
elements is less than 1 yr (Koester 2009), and it is straightforward to determine the true abundances in the parent body from measurements of the
the relative photospheric abundances of the different elements and theoretical estimates of their settling times.    Therefore, with measures
of oxygen, silicon, iron and magnesium for externally-polluted hydrogen white dwarfs, it should  be possible to determine if there is ``excess" oxygen.  While ground-based optical observations have not yielded this suite of abundances in hydrogen white dwarfs, high quality ultraviolet data obtained from
space should provide the needed measurements. For example, with 6 {\AA} resolution, Koester, Provencal \& Shipman (1997) reported Mg and Fe
in G 29-38, a hydrogen white dwarf with substantial pollution by heavy elements.   With higher spectral resolutions  now available,  many more elements including oxygen could be detected in the atmosphere of this and other similar externally polluted stars.     If ``excess" oxygen is found,
the system is a strong candidate for having accreted water.

If H$_{2}$O is common
in extrasolar minor planets, then likely  there would be many environments with liquid water.  By analogy with Solar System objects, there could be extrasolar minor
planets that are large enough that internal heating is important.  Consequently, as with Europa,  there could be an internal ocean (Kivelson et al. 2000).
Also, as in the
early Solar System, asteroids could deliver water to analogs of the Earth which could then develop their own surface oceans.    Analysis of white dwarf pollutions provides a pathway for  investigating the hypothesis that environments with
liquid water that are inhabitable by extremophiles are widespread in the Milky Way.

This work has been partly supported by NASA and the NSF.  We thank D. Koester for generously providing model calculations.

\newpage
\begin{center}
Table 1  -- White Dwarf Pollutions
\\
\begin{tabular}{llcllcl}
Z & &GD 61 & && GD 378 & \\
\hline
&  $M_{mix}(Z)^{a}$ & $t_{set}^{b}$ & ${\dot M}_{*}(Z)^{c}$ & $M_{mix}(Z)^{a}$ & $t_{set}^{b}$ &${\dot M}_{*}(Z)^{c}$\\
 & (10$^{20}$ g)& (10$^{5}$ yr) & (10$^{8}$ g s$^{-1}$) & (10$^{20}$ g) & (10$^{5}$ yr)& (10$^{8}$ g s$^{-1}$) \\
 \hline
 \hline
 H & 55 & ... & ... & 90 & ... & ... \\
  C & $<$0.010 & 1.22 & $<$0.0026  &2.4& 3.5 & 0.22 \\
 O & 13 & 0.92 & 4.5 & 26 &3.0 & 2.8 \\
 Si & 2.9 &  0.64 &1.4 &  2.8 & 2.1& 0.42 \\
 S & $<$0.21 & 0.54 & $<$0.12& 1.0 & 1.7 & 0.19 \\
  Ca & 0.66 & 0.49 & 0.43 &  0.81 & 1.5 & 0.17 \\
 Fe & 0.74 & 0.35 & 0.67 & 4.5& 1.1& 1.3 \\

\hline
 \end{tabular}
 \end{center}
$^{a}$Masses of the elements in the mixing zone are derived from the  number abundances given by  Desharnais et al. (2007) and the total mass of the mixing zone from  Koester (private communication) who found values for
 GD 61 and GD 378 of 2.1 ${\times}$ 10$^{26}$ and 6.4 ${\times}$ 10$^{26}$ g, respectively.  The calculations for these individual stars  are described based on the models described by  Koester (2009) with specific atmospheric parameters from Desharnais et al. (2009.  For GD 61,  $T$ = 17,280 K, $\log$ $g$ = 8.20 [cgs units] and $\log$ $m$(H)/$m$(He) = -4.58, while for GD 378,
 $T$ = 16,600, $\log$ $g$ = 8.03 [cgs units] and $\log$ $m$(H)/$m$(He) = - 4.85.
 \\
 $^{b}$from Koester (private communication) based on Koester (2009)
 \\
 $^{c}{\dot M}_{mix}(Z)$ is defined  as $M_{mix}(Z)/t_{set}(Z)$ (Jura et al. 2009)
  \newpage
\begin{figure}
%\plotone{fig1.pdf}
\plotone{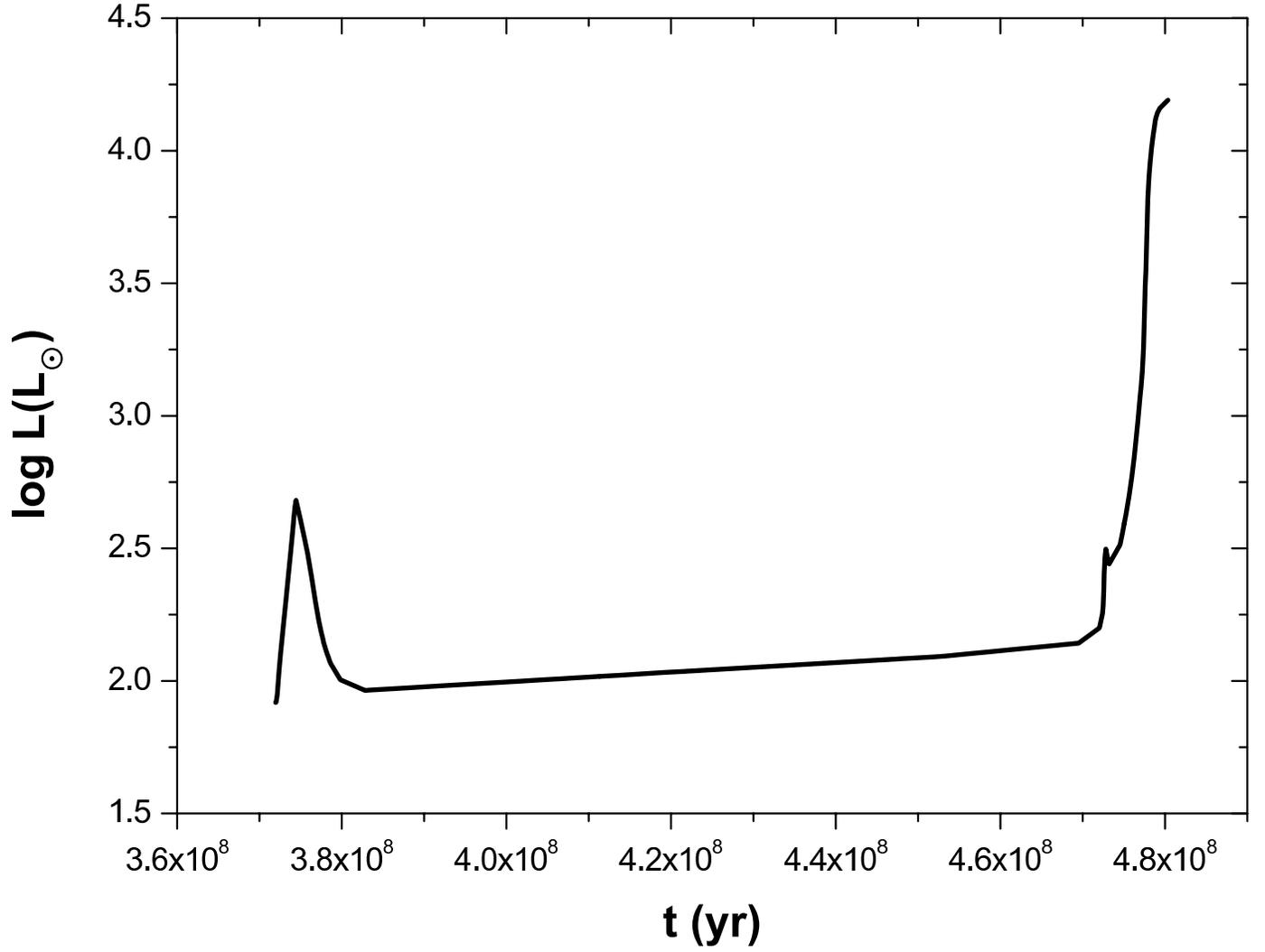}
\caption{Luminosity vs. time for a 3 M$_{\odot}$ star (Girardi et al. 2000).   The first spike in $L$ at $t$ ${\sim}$ 400 Myr is the first ascent up the red
giant branch while the second spike represents the AGB phase of evolution.  }
\end{figure}
\newpage
\begin{figure}
%\plotone{fig2.pdf}
\plotone{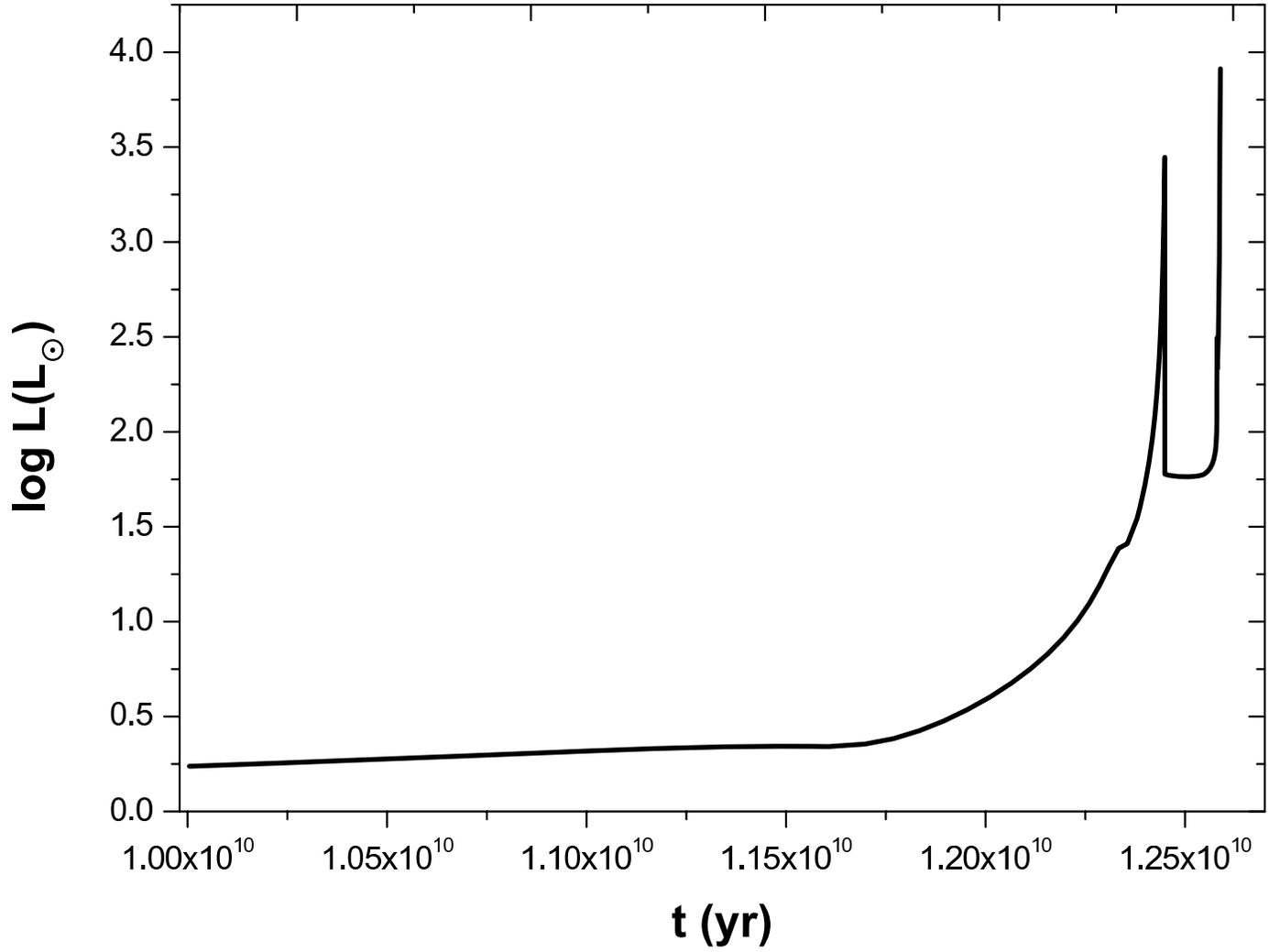}
\caption{Similar to Fig. 1 except for a 1 M$_{\odot}$ star and using the calculations from Pols et al. (1998).}
\end{figure}
\newpage
\begin{figure}
%\plotone{fig3.pdf}
\plotone{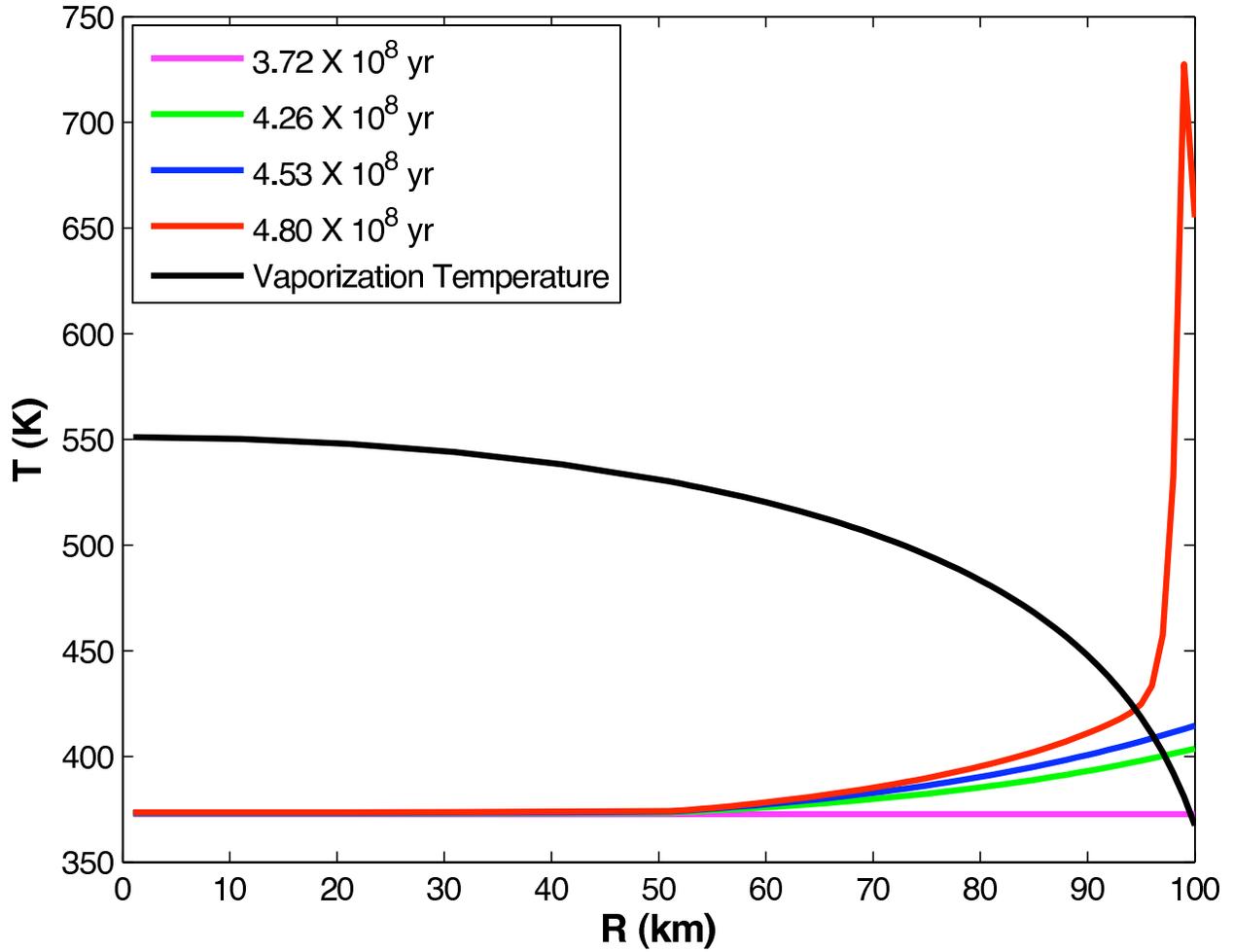}
\caption{Internal temperature profiles at different times for a 100 km radius minor planet at an initial orbital radius around of 5 AU around a star with a main-sequence mass of 3 M$_{\odot}$.  The red curve denotes the final thermal profile in the interior of the minor planet before the star leaves the AGB on its evolutionary path to becoming a white dwarf.}
\end{figure}
\newpage
\begin{figure}
%\plotone{fig4.pdf}
\plotone{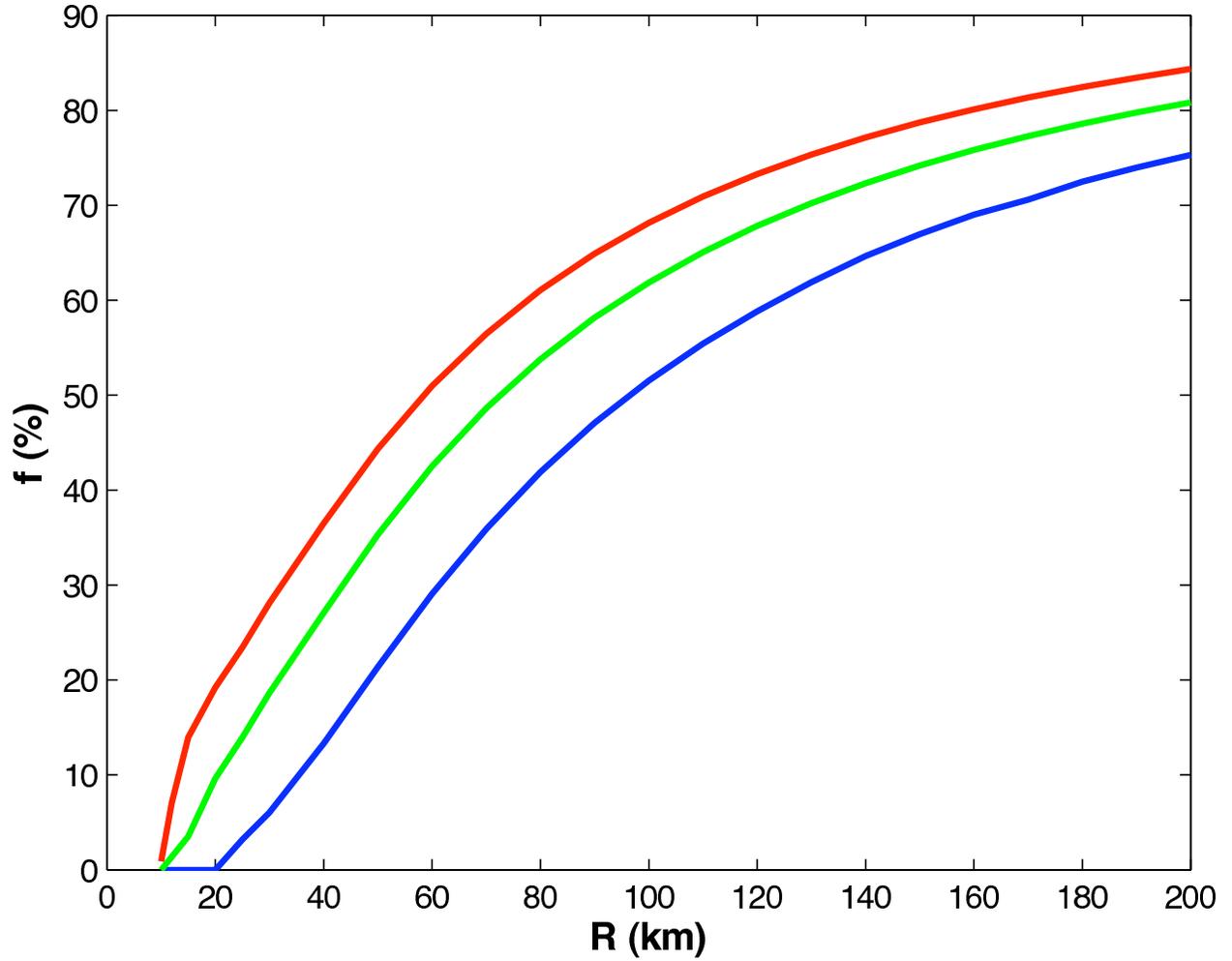}
\caption{Mass percentage of retained ice ($f$) at the end of the AGB evolution for minor planets of radius $R$ for a star of main-seqeunce mass 3 M$_{\odot}$.  The
blue, green and red curves refer to asteroids of initial orbital radii of 5 AU, 7.5 AU and 10 AU, respectively.}
\end{figure}
\newpage
\begin{figure}
%\plotone{fig5.pdf}
\plotone{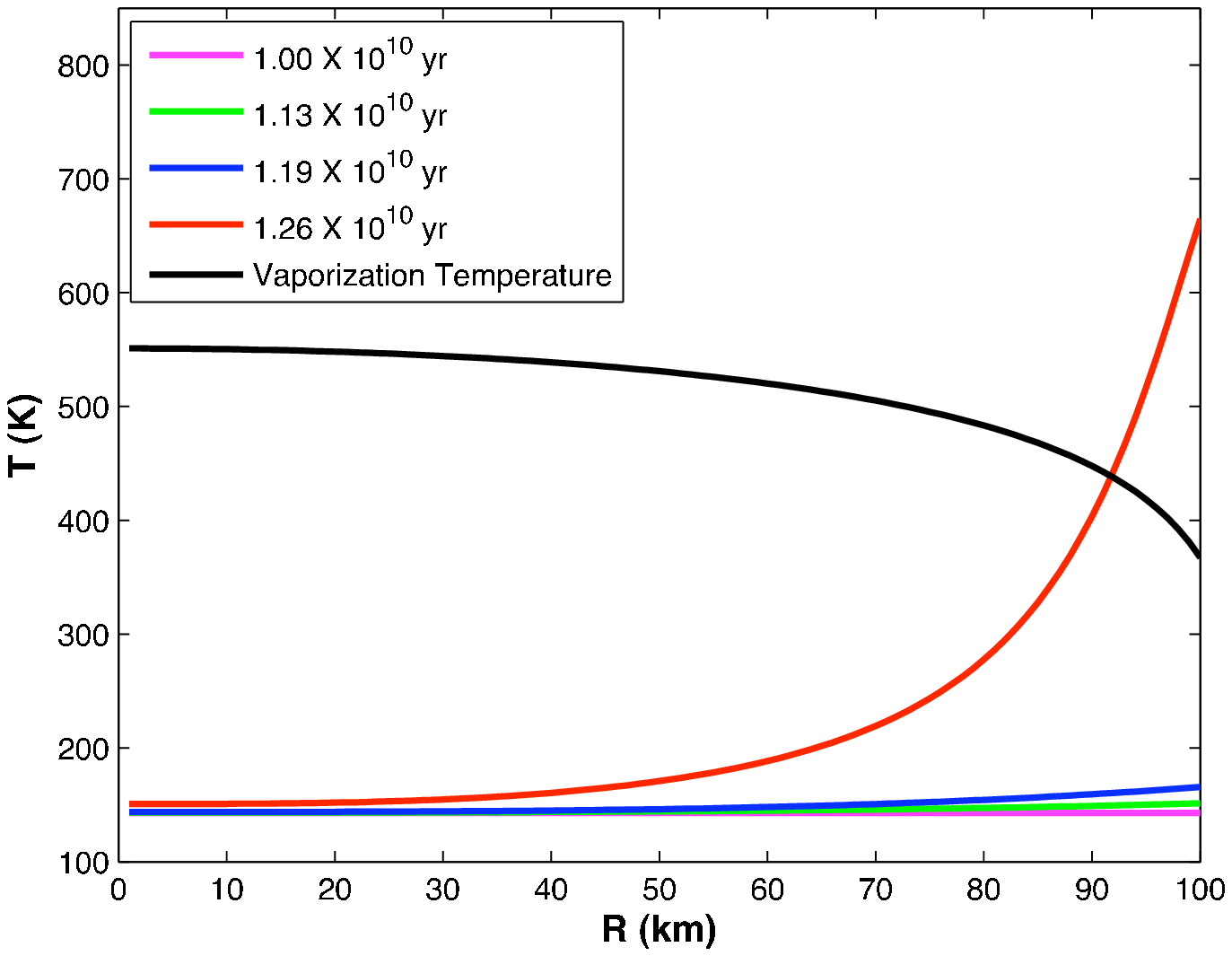}
\caption{Same as Fig. 3 except for a star with a main-sequence mass of 1 M$_{\odot}$.}
\end{figure}
\newpage
\begin{figure}
%\plotone{fig6.pdf}
\plotone{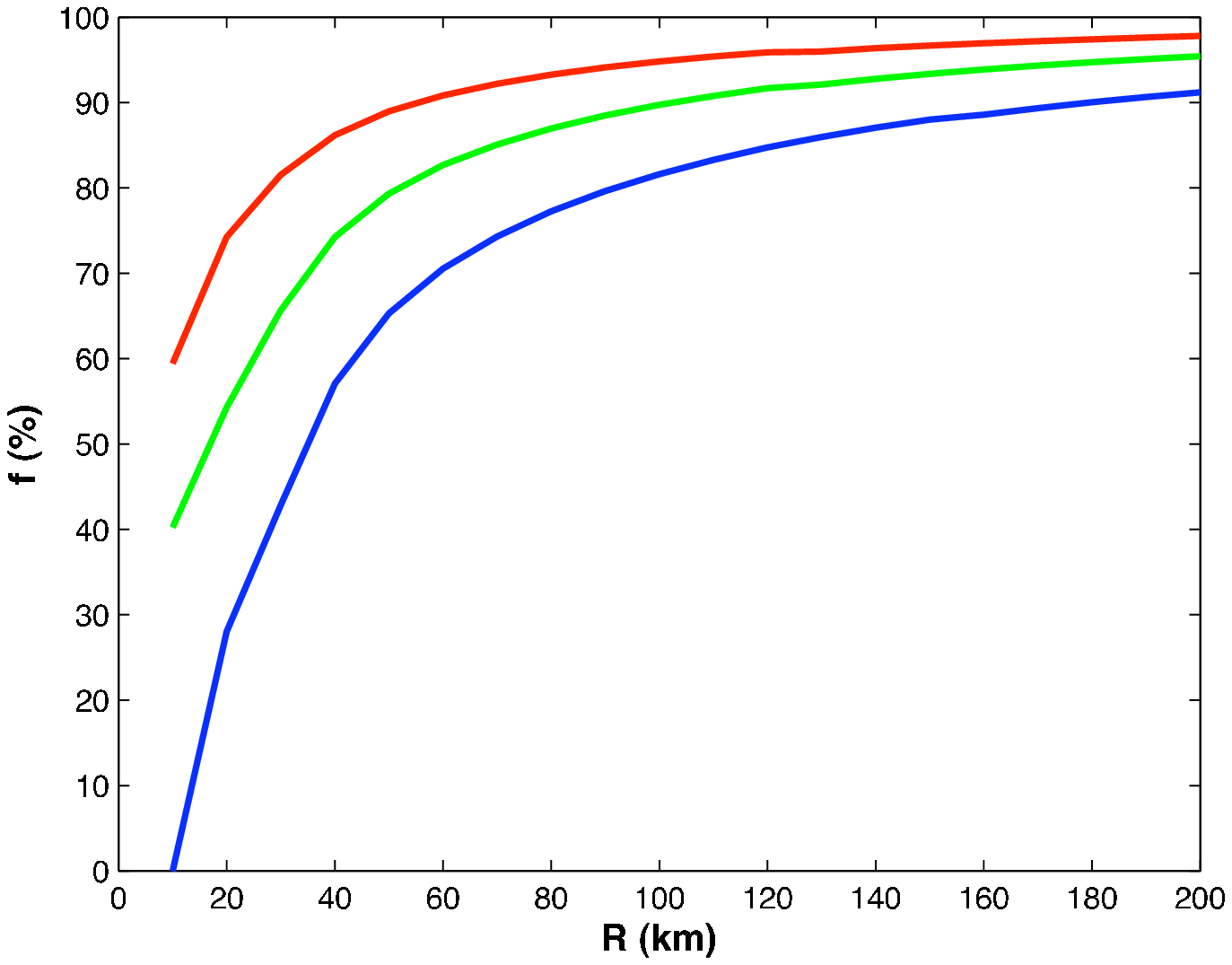}
\caption{Same as Fig. 4 except for a star with a main sequence mass of 1 M$_{\odot}$.}
\end{figure}
\newpage
\begin{figure}
%\plotone{fig7.pdf}
\plotone{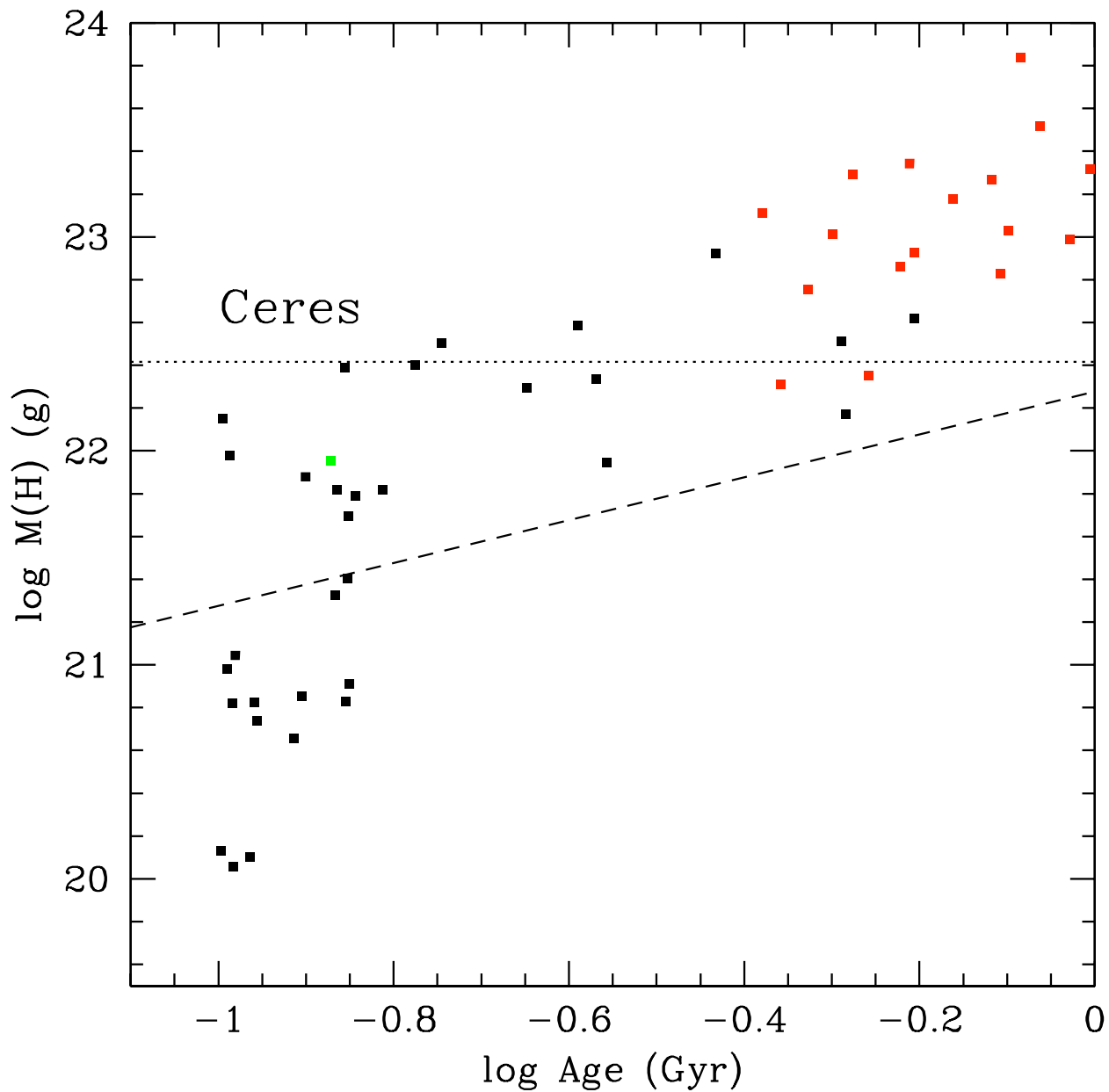}
\caption{Mass of hydrogen in the mixing zone vs. cooling age for helium white dwarfs. The black and red squares show data from Voss et al. (2007) and Dufour et al. (2007), respectively while the green square represents GD 378.   Upper limits reported in these papers are not displayed; therefore this plot represents the stars which
have higher than average amounts of atmospheric hydrogen.  The dotted horizontal line shows the estimated mass of internal hydrogen for Ceres, the largest asteroid in the Solar System. The sloping dashed line shows the expected amount of mass for
an accretion rate of 6 ${\times}$10$^{5}$ g s$^{-1}$. }
\end{figure}

\end{document}